\documentstyle{article}

\font\tenrm=cmr10

\font\elevenbf=cmbx10 scaled\magstep 1
\font\elevenrm=cmr10 scaled\magstep 1
\font\elevenit=cmti10 scaled\magstep 1

\renewenvironment{thebibliography}[1]
 { \elevenrm
   \begin{list}{\arabic{enumi}.}
    {\usecounter{enumi} \setlength{\parsep}{0pt}
     \setlength{\itemsep}{3pt} \settowidth{\labelwidth}{#1.}
     \sloppy
    }}{\end{list}}

\begin{document}

\baselineskip=13pt

\rightline {\bf UG-FT-35/94}

\begin{center}
\vglue 1.5cm
{
 {\elevenbf        \vglue 10pt
               The Physics of ${\bf Z'}$ bosons
               \footnote{
\tenrm
\baselineskip=11pt
Lectures delivered at the XVII International School of
Theoretical Physics "Standard Model $\&$ Beyond '93",
Szczyrk, September 19-27, 1993\\}
\vglue 5pt}

\vglue 1.5cm
{\tenrm F. DEL AGUILA \\}
\baselineskip=13pt
\vglue 0.2cm
{\elevenit Departamento de F\'\i sica Te\'orica y del Cosmos,
Universidad de Granada \\}
\baselineskip=12pt
{\elevenit Granada, 18071, Spain \\}}

\vglue 1.5cm
{\tenrm ABSTRACT}

\end{center}

\vglue 0.3cm
{\rightskip=3pc
 \leftskip=3pc
 \tenrm\baselineskip=12pt
 \noindent
In these lectures we review the simplest gauge extensions
of the standard model, and the present and future limits
on new weak interactions.}

\vglue 1.cm
{\centerline{\elevenbf 1. Introduction}}
\vglue 0.5cm

\elevenrm
The physics of $Z'$ bosons addresses two main questions:
Which is the reason for extra weak interactions ? How
does a new neutral gauge boson manifest at the
electroweak scale ? The first question is a model
building one and it is believed to be related with the
physics at the Planck scale. The second question is a
phenomenological one and it will be our main interest
in these lectures.

The standard model gives a description of Nature below
$100\ GeV$ in impressive agreement with experiment [1].
As a result present data put non-trivial bounds on
possible extensions of the standard model, in particular
on any electroweak gauge extension. However, these limits
are very model dependent and rather weak for some models.
There is a large literature on the subject to which we
refer in the text. For more detailed lectures on $Z'$
bounds from LEP see T. Riemann at this School [2].

We organize these lectures revising first the simplest
($E_6$) gauge extensions of the standard model. In a
short version of the second lecture we summarize present
indirect limits on new gauge interactions. Whereas the
third lecture is devoted to
recent developments on the diagnostics of $Z'$ gauge
couplings to known quarks and leptons at large hadron
and lepton colliders.
\eject

\vglue 1.cm
{\centerline{\elevenbf 2. Simplest (${\bf E_6}$)
gauge extensions of the standard model}}
\vglue 0.5cm

The purpose of this lecture is to introduce a general,
convenient parametrization of (minimal) extended
electroweak models in order to compare with experiment.

The simplest gauge extension of the standard model results
from adding only one extra neutral gauge boson, $Z'$.
The corresponding gauge group enlarges the
electroweak standard model by including one extra
$U'(1)$ factor: $SU(2)_L\times U(1)_Y\times U'(1)$.
In this case and in what follows we will assume that
the new charge $Q'$ commutes with the $SU(2)_L$
generators $T_i$: $[Q',T_i] = 0$. We will
also assume family universality.

Prior to the introduction of popular ($E_6$) gauge
extensions of the standard model, let us discuss the
general parametrization of the gauge interactions for a
gauge group with two abelian $U(1)$ factors [3].

\vskip 0.5cm
{\centerline
{\elevenit 2.1 General tree level lagrangian for $U^1(1)
\times U^2(1)$}}
\vskip 0.5cm

For the sake of definiteness we consider $n$ massless
Weyl fermions $\psi _k, k=1,...,n,$ with different
$U^a(1), a=1,2,$ charges $Y^a_k$, and then unmixed.
Calling $A_{\mu}^b, b=1,2,$ the massless gauge bosons,
the most general lagrangian reads

\begin{equation}
{\cal L}=-{\frac{1}{4}}F^b_{\mu \nu}F^{b \mu \nu}
+\bar{\psi}_ki{\not\partial}\psi _k
+\bar{\psi}_k\gamma ^{\mu}Y^a_k\psi _kg_{ab}A_{\mu}^b
\end{equation}

\noindent
where summation on repeated indices is understood. This
would be the standard QED lagrangian but for two different
gauge interactions if $g_{ab}$ were diagonal. However,
$g_{ab}$ is not diagonal in general. There is no symmetry
reason for being so, and, moreover, the off-diagonal gauge
couplings are observable !

\begin{itemize}
\item Looking for known examples, the neutral sector of the
standard model $U(1)_{T_3}\times U(1)_Y$ is {\elevenit
not general enough}. In this case $g_{ab}$ is diagonal ! But
there is a symmetry reason: {\elevenit gauge invariance}.
$U(1)_{T_3}$ is within $SU(2)_L$ and no mixing among the
non-abelian $SU(2)_L$ and the abelian $U(1)_Y$ massless
gauge bosons is allowed.
\item Usually when illustrating the
electroweak gauge extensions
of the standard model by adding an extra $U'(1)$, the possible
mixing with the standard $U(1)_Y$ factor is {\elevenit
not discussed}. This is only {\elevenit justified under
certain assumptions}. At any rate, in the particular cases
usually discussed the numerical effects of this mixing
are {\elevenit small}.
\item These off-diagonal gauge couplings are necessary
when working to higher orders in perturbation theory
for quantum corrections {\elevenit bring them back}.
They are also necessary to renormalize the theory because
{\elevenit a full set of renormalization conditions} is
required.
\item This mixing allows for a general parametrization of
an extra neutral gauge interaction and for
the possibility of
{\elevenit fitting} to precise data {\elevenit minimazing
within the corresponding class of models}. This class
includes the popular $E_6$ models used often
for illustration.
\end{itemize}

Using the invariance of the previous lagrangian under the
rotation of the gauge bosons we can always assume that the
matrix $g_{ab}$ is triangular, $g_{ab}=0, a>b$. The
resulting $\frac{1}{2}N(N+1)$ ($=3$ for $N=2$) couplings
become then physical. Once a basis for the charges
$Y^a_k$ defining $U^1(1)\times U^2(1)$ is chosen, the
fixed triangular matrix $g_{ab}$ can be measured,
and it is not correct in general to assume
it to be diagonal.

\vskip 0.5cm
{\centerline
{\elevenit 2.2 General tree level parametrization for
$SU(2)_L\times U(1)_Y\times U'(1)$}}
\vskip 0.5cm

This parametrization was first introduced in Ref. [4].
The explicit form discussed below was presented in detail
in Ref. [5]. The interacting lagrangian in Eq. (1) for
$SU(2)_L\times U(1)_Y\times U'(1)$ and the standard model
fermions reads:

\begin{equation}
-{\cal L}_{NC}=
\bar{\psi}_k \gamma ^{\mu} \{
T_{3k}g W_{3\mu} + Y_kg'_{11} B'_{\mu}
+ Y_kg'_{12} Z'_{2\mu} + Q'_kg'_{21} B'_{\mu}
+ Q'_kg'_{22} Z'_{2\mu}
\} \psi _k
\end{equation}

\noindent
where $T_{3}$ and $Y$ are the standard isospin and
hypercharge charges (see Table 1) and $Q'$ is the extra
$U'(1)$ charge. As emphasized above $g'_{ab}$ can be made
triangular by rotating the abelian gauge bosons:

\begin{equation}
-{\cal L}_{NC}=
\bar{\psi}_k \gamma ^{\mu} \{
T_{3k}g W_{3\mu} + Y_kg_{11} B_{\mu}
+ Y_kg_{12} Z_{2\mu}
+ Q'_kg_{22} Z_{2\mu}
\} \psi _k
\end{equation}

\noindent
where $B'_{\mu}=c'B_{\mu}-s'Z_{2\mu},
Z'_{2\mu}=s'B_{\mu}+c'Z_{2\mu}$. $c'$ and $s'$
are the cosinus and sinus of the corresponding rotation
angle fulfilling $g_{21}=g'_{21}c'+g'_{22}s'=0$; whereas
$g_{11}=g'_{11}c'+g'_{12}s', g_{12}=-g'_{11}s'+g'_{12}c',
g_{22}=-g'_{21}s'+g'_{22}c'$ are {\elevenit measurable}
couplings. The photon is defined as a combination of
$W_{3\mu}$ and $B_{\mu}$:
$W_{3\mu}=s_WA_{\mu}+c_WZ_{1\mu},
B_{\mu}=c_WA_{\mu}-s_WZ_{1\mu}$.
The extended neutral current lagrangian for the standard
Dirac fermions can then be written in general [6]:

\begin{equation}
-{\cal L}_{NC}=
e \sum _{i} q_i \bar{\psi}_i \gamma ^{\mu} \psi _i A_{\mu}
+ \frac {g}{2c_W} \sum _{i} \bar{\psi}_i \gamma ^{\mu}
(v^i - a^i\gamma _5) \psi _i Z_{1\mu}
+ \frac {g}{2c_W} \sum _{i} \bar{\psi}_i \gamma ^{\mu}
(v'^i - a'^i\gamma _5) \psi _i Z_{2\mu}
\end{equation}

\noindent
with the charges given in Table 2,
$e=\frac {\sqrt  {\frac {3}{5}}gg_{11}}
{\sqrt{g^2+\frac {3}{5}g_{11}^2}}$
is the positron charge, and $s_W \equiv sin\theta _W =
\frac {e}{g}$ is the electroweak mixing angle ($c_W =
cos\theta _W$). Hence any extra neutral gauge
interaction can be written in the current eigenstate
gauge boson basis adding an extra term
for the new gauge boson similar to the standard model
one but allowing for a linear
combination of the extra charge $Q'$ and the standard
model hypercharge $Y$: $g_{22}Q'+g_{12}Y$ in Eq. (3).
Once the charge of the extra $U'(1)$, $Q'$,
is specified, $v'^i$ and $a'^i$
in Eq. (4) can be calculated (see below for popular
models). They are functions of $g_2\equiv g_{22}$ and
$\frac {g_{12}}{g_2}$. However, $g_2$
is often the only coupling strength introduced in the
literature for the
new interaction. It is also often taken equal or similar
to $g_{11}=\sqrt{\frac {5}{3}}\frac {e}{c_W}$. But in
general the neutral current lagrangian for
$SU(2)_L\times U(1)_Y\times U'(1)$ depends on $4$
independent, measurable parameters:
$e, s_W, g_2$ and $\frac {g_{12}}{g_2}$.

\vglue 0.2cm
\begin{center}
\begin{tabular}{|c|ccc|}
\hline
\multicolumn{1}{|c}{} & $T_{3}$
& $\sqrt {\frac {5}{3}}Y$ &
\multicolumn{1}{c|}{$Q'$} \\
\hline
$ \left(\begin{array}{c}
u \\ d \\ \end{array}\right) _L $
& $ \begin{array}{c}
\frac {1}{2} \\ -\frac {1}{2} \\ \end{array}\} $
& $ \frac {1}{6} $ &
$ q'_{q_L} $ \\
$u_L^c$ & $ 0 $ & $ -\frac {2}{3} $ &
$ q'_{u^c_L} $ \\
$d_L^c$ & $ 0 $ & $ \frac {1}{3} $ &
$ q'_{d^c_L} $ \\
$ \left(\begin{array}{c}
\nu \\ e \\ \end{array}\right) _L $
& $ \begin{array}{c}
\frac {1}{2} \\ -\frac {1}{2} \\ \end{array}\} $
& $ -\frac {1}{2} $ &
$ q'_{l_L} $ \\
$e_L^c$ & $ 0 $ & $ 1 $ &
$ q'_{e^c_L} $ \\
\hline
\end{tabular}
\end{center}
\vglue 0.1cm

\vglue 0.1cm
\rightskip=3pc
\leftskip=3pc
{\tenrm \baselineskip=12pt
\noindent
Table 1. Isospin $T_3$, hypercharge $Y$, and new charge
$Q'$ assignments for ordinary fermions.}
\vglue 0.1cm

\rightskip=0pc
\leftskip=0pc
\vglue 0.2cm
\begin{center}
\begin{tabular}{|c|ccccc|}
\hline
\multicolumn{1}{|c}{} & $Q$
& $V$ & $A$ & $\frac {g}{c_W}V'$ &
\multicolumn{1}{c|}{$\frac {g}{c_W}A'$} \\
\hline
$ u $ & $ \frac {2}{3} $ &
$ \frac {1}{2}-\frac {4}{3} sin^2\theta _W$ &
$ \frac {1}{2} $ &
$g_2[(q'_{q_L}-q'_{u^c_L})+
\frac {g_{12}}{g_2} \sqrt {\frac {3}{5}}\frac {5}{6}] $ &
$g_2[(q'_{q_L}+q'_{u^c_L})+
\frac {g_{12}}{g_2} \sqrt {\frac {3}{5}}(-\frac {1}{2})] $ \\
$ d $ & $ -\frac {1}{3} $ &
$ -\frac {1}{2}+\frac {2}{3} sin^2\theta _W $ &
$ -\frac {1}{2} $ &
$g_2[(q'_{q_L}-q'_{d^c_L})+
\frac {g_{12}}{g_2} \sqrt {\frac {3}{5}}(-\frac {1}{6})] $ &
$g_2[(q'_{q_L}+q'_{d^c_L})+
\frac {g_{12}}{g_2} \sqrt {\frac {3}{5}}\frac {1}{2}] $ \\
$ \nu $ & $ 0 $ &
$ \frac {1}{2} $ &
$ \frac {1}{2} $ &
$g_2[q'_{l_L}+
\frac {g_{12}}{g_2} \sqrt {\frac {3}{5}}(-\frac {1}{2})] $ &
$g_2[q'_{l_L}+
\frac {g_{12}}{g_2} \sqrt {\frac {3}{5}}(-\frac {1}{2})] $ \\
$ e $ & $ -1 $ &
$ -\frac {1}{2} +2 sin^2\theta _W $ &
$ -\frac {1}{2} $ &
$g_2[(q'_{l_L}-q'_{e^c_L})+
\frac {g_{12}}{g_2} \sqrt {\frac {3}{5}}(-\frac {3}{2})] $ &
$g_2[(q'_{l_L}+q'_{e^c_L})+
\frac {g_{12}}{g_2} \sqrt {\frac {3}{5}}\frac {1}{2}] $ \\
\hline
\end{tabular}
\end{center}
\vglue 0.1cm

\vglue 0.1cm
\rightskip=3pc
\leftskip=3pc
{\tenrm \baselineskip=12pt
\noindent
Table 2. Electric charge, $Q=T_3+\sqrt {\frac {5}{3}}Y$,
and vector and axial couplings for $Z_{1\mu}$:
$v^i=t_{3i}-2q_isin^2\theta _W, a^i=t_{3i}$,
and $Z_{2\mu}$: $\frac {g}{c_W}v'^i=g_2[(q'_{iL}+q'_{iR})+
\frac {g_{12}}{g_2} \sqrt {\frac {3}{5}}(-t_{3i}+2q_i)],
\frac {g}{c_W}a'^i=g_2[(q'_{iL}-q'_{iR})+
\frac {g_{12}}{g_2} \sqrt {\frac {3}{5}}(-t_{3i})]$,
for ordinary fermions in Table 1.}
\vglue 0.1cm

\rightskip=0pc
\leftskip=0pc
\vskip 0.5cm
{\centerline
{\elevenit 2.3 $SO(10), E_6$, superstring-motivated and $LR$
models}}
\vskip 0.5cm

Popular extended electroweak models are particular cases
of Eq. (4) and Table 2:

\begin{itemize}
\item model $\chi$ occurs when $SO(10)$ breaks down to
$SU(5)\times U(1)_{\chi}$;
\item model $\psi$ occurs when $E_6$ breaks down to
$SO(10)\times U(1)_{\psi}$; and
\item model $\eta$ occurs in superstring-inspired models
in which $E_6$ breaks directly to a rank 5 group [7].
\end{itemize}

\noindent
The corresponding charges are given in Table 3.

\vglue 0.2cm
\begin{center}
\begin{tabular}{|c|ccc|}
\hline
\multicolumn{1}{|c}{$Q'$} & $2\sqrt {10}Q_{\chi}$
& $2\sqrt {6}Q_{\psi}$ &
\multicolumn{1}{c|}{$2\sqrt {15}Q_{\eta}$} \\
\hline
$ q'_{q_L} $
& $ -1 $ & $ 1 $ & $ -2 $ \\
$ q'_{u^c_L} $ & $ -1 $ & $ 1 $ & $ -2 $ \\
$ q'_{d_L^c} $ & $ 3 $ & $ 1 $ & $ 1 $ \\
$ q'_{l_L} $
& $ 3 $ & $ 1 $ & $ 1 $ \\
$ q'_{e_L^c} $ & $ -1 $ & $ 1 $ & $ -2 $ \\
\hline
\end{tabular}
\end{center}
\vglue 0.1cm

\vglue 0.1cm
\rightskip=3pc
\leftskip=3pc
{\tenrm \baselineskip=12pt
\noindent
Table 3. Popular $Z'$ model charges.}
\vglue 0.1cm

\rightskip=0pc
\leftskip=0pc
\vglue 0.1cm
\begin{itemize}
\item The general $E_6$ extra $U(1)$ depends on a mixing
angle, $\beta$, which fixes the combination of
two extra, independent $U(1)$'s arbitrarily chosen
within $E_6$, {\elevenit
e.g.}, $U(1)_{\chi}$ and $U(1)_{\psi}$:
$Q_{E_6}=cos\beta Q_{\chi}+sin\beta Q_{\psi}$.
\end{itemize}

\noindent
$\chi , \psi$ and $\eta$ correspond to $\beta =
0, \frac {\pi}{2}, -arctan \sqrt{\frac {5}{3}}$,
respectively.
In all these models $g_2 = \sqrt {\frac {5}{3}}\frac {e}
{c_W}, g_{12} = 0$. The neutral current lagrangian for
left-right ($LR$)
models is also a particular case of Eq. (4) and Table 2.

\begin{itemize}
\item $LR$ models are parametrized by the ratio $\kappa =
\frac {g_R}{g_L}$ of the gauge couplings $g_{L,R}$ for
$SU(2)_{L,R}$, respectively. $\kappa > \frac {s_W}{c_W}$
for consistency [8].
\end{itemize}

\noindent
The extra charge $Q_{LR}=Q_{\chi}$, because it is the
only extra one within $SO(10)$. In this case, however,
$g_2 = \frac {e}{c_W}\sqrt {\frac {2}{5}}
\frac {\beta ^2+1}{\beta}, \frac {g_{12}}{g_2} =
\sqrt {\frac {1}{6}}\frac {3\beta ^2-2}
{\beta ^2+1}, \beta = \sqrt {\kappa ^2
cot^2\theta _W-1}$.
Although $g_2$ and $\frac {g_{12}}{g_2}$ are more
convenient for comparison with experiment, unification
of gauge couplings at a very high scale motivates an
alternative parametrization [5]:

\begin{equation}
g_2=\sqrt {\frac {5}{3}}\frac {e}{c_W}
(\frac {c_1^2}{\lambda}+\lambda s_1^2),
\frac {g_{12}}{g_2} =
\frac {s_1c_1(\lambda ^2-1)}{s_1^2\lambda ^2+c_1^2},
\end{equation}

\noindent
where $s_1(c_1)\equiv sin\theta _1 (cos\theta _1),
\theta _1 \in [-\frac {\pi}{2}, \frac {\pi}{2}),
\lambda \geq 1$. $\beta $ is also replaced by $\theta _2 =
-\beta - arctan \sqrt {\frac {5}{3}}$.
(In any case care must be
taken with the different ranges of variation
and sign conventions existing in the literature.)

\vskip 0.5cm
{\centerline
{\elevenit 2.4 $Z'$ lagrangian parametrization}}
\vskip 0.5cm

In extended electroweak models heavy gauge bosons can
mix. Thus,

\begin{equation}
Z_{1\mu}=c_3Z_{\mu}+s_3Z'_{\mu},
Z_{2\mu}=-s_3Z_{\mu}+c_3Z'_{\mu}
\end{equation}

\noindent
in Eq. (4). $Z$ is the observed heavy neutral gauge boson
and $Z'$ is the new gauge boson; whereas
$s_3(c_3)\equiv sin\theta _3 (cos\theta _3)$ is the
$Z_1Z'$ mixing.

\vskip 0.5cm
{\centerline
{\elevenit 2.5 Relevant couplings for large hadron and lepton
colliders}}
\vskip 0.5cm

We can look for indirect $Z'$ effects in precise
electroweak data and for direct $Z'$ production at large
colliders. Precise low energy
and in particular LEP data test mainly the $Z_{\mu}$
current (see Eqs. (4, 6)):

\begin{equation}
\frac {g}{c_W} J^{\mu} =
\frac {g}{2c_W} \sum _{i} \bar{\psi}_i \gamma ^{\mu}
[ (v^ic_3 -v'^is_3) - (a^ic_3-a'^is_3)\gamma _5 ] \psi _i.
\end{equation}

\noindent
As no indication for new effects beyond the standard model
is found, it is deduced that only a small gauge boson
mixing is allowed. (See Ref. [9] for a recent discussion.
See also T. Riemann lectures at this School [2].)  Present
limits require $s_3 < 1\%$.

For $Z'$ physics at future colliders more convenient sets
of effective charges can be defined [10, 11].
Production bounds test mainly the $Z'_{\mu}$ current:

\begin{equation}
\frac {g}{c_W} J'^{\mu} =
\frac {g}{2c_W} \sum _{i} \bar{\psi}_i \gamma ^{\mu}
[ (v'^ic_3 +v^is_3) - (a'^ic_3+a^is_3)\gamma _5 ] \psi _i.
\end{equation}

\noindent
The precision available at large colliders and the
experimental limit on $s_3$ permit to neglect the
$Z_1Z'$ mixing. The $Z'$ couplings to ordinary
fermions is fixed by five
normalized charges $\hat g^u_{L2} = \hat g^d_{L2} \equiv
\hat g^q_{L2}, \hat g^u_{R2}, \hat g^d_{R2},
\hat g^{\nu}_{L2} = \hat g^e_{L2} \equiv \hat g^l_{L2},
\hat g^e_{R2}$ and the gauge coupling strength
$g_2$:

\begin{equation}
g_2 \hat g^i_{L,R2}\equiv \frac {g}{2c_W}(v'^i\pm a'^i).
\end{equation}

\noindent
The signs of the charges will be hard to determine at
hadron colliders. The following set of four normalized
couplings is probed directly [10]:

\begin{equation}
\gamma ^l_L\equiv \frac {(\hat g^l_{L2})^2}
{(\hat g^l_{L2})^2+(\hat g^e_{R2})^2},
\gamma ^q_L\equiv \frac {(\hat g^q_{L2})^2}
{(\hat g^l_{L2})^2+(\hat g^e_{R2})^2},
\tilde U\equiv \frac {(\hat g^u_{R2})^2}
{(\hat g^q_{L2})^2},
\tilde D\equiv \frac {(\hat g^d_{R2})^2}
{(\hat g^q_{L2})^2}.
\end{equation}

\noindent
At large lepton colliders the main effects of a new
gauge boson far off-shell result from its interference
with the photon and the $Z$ boson. Since the photon
couplings are only vector-like and the $l$ couplings to
$Z$ are almost axial,
it turns out that the probes in the two-fermion final
state channels single out the $Z'$ leptonic couplings
primarily in the combinations
$\hat g^l_{L2}\pm \hat g^e_{R2}$. To trace the
combinations of the normalized charges to which the probes
are sensitive, it is advantageous to choose either of the
two combinations to normalize the charges. We choose the
$\hat g^l_{L2}-\hat g^e_{R2}$ combination, which turns out
to be a convenient choice for popular models. We then
define the following four independent normalized charges
[11]:

\begin{equation}
P_V^e = \frac {\hat g^l_{L2} + \hat g^e_{R2}}
{\hat g^l_{L2} - \hat g^e_{R2}},
\ P_L^q = \frac {\hat g^q_{L2}}
{\hat g^l_{L2} - \hat g^e_{R2}},
\ P_R^{u,d} = \frac {\hat g^{u,d}_{R2}}
{\hat g^q_{L2}}.
\end{equation}

\noindent
In Table 4 we give the values of the couplings in Eqs.
(10, 11) for models $\chi , \psi , \eta $ and $LR$ for
the special value $\kappa = 1$.

\vglue 0.2cm
\begin{center}
\begin{tabular}{|c|cccc|}
\hline
\multicolumn{1}{|c}{} & $\chi$
& $\psi$ & $\eta$ &
\multicolumn{1}{c|}{$LR$} \\
\hline
$\gamma ^l_L$ & $ 0.9 $ & $ 0.5 $ &
$ 0.2 $ & $ 0.36 $ \\
$\gamma ^q_L$ & $ 0.1 $ & $ 0.5 $ &
$ 0.8 $ & $ 0.04 $ \\
$\tilde U$ & $ 1 $ & $ 1 $ &
$ 1 $ & $ 37 $ \\
$\tilde D$ & $ 9 $ & $ 1 $ &
$ 0.25 $ & $ 65 $ \\
& & & & \\
$ P_V^e $ & $ 2 $ & $ 0 $ &
$ -3 $ & $ -0.148 $ \\
$ P_L^q $ & $ -0.5 $ & $ 0.5 $ &
$ 2 $ & $ -0.142 $ \\
$ P_R^u $ & $ -1 $ & $ -1 $ &
$ -1 $ & $ -6.04 $ \\
$ P_R^d $ & $ 3 $ & $ -1 $ &
$ 0.5 $ & $ 8.04 $ \\
\hline
\end{tabular}
\end{center}
\vglue 0.1cm

\vglue 0.1cm
\rightskip=3pc
\leftskip=3pc
{\tenrm \baselineskip=12pt
\noindent
Table 4. Couplings in Eqs. (10, 11) for models $\chi ,
\psi , \eta $ and $LR$ (with $\kappa = 1$
and $s^2_W = 0.23$).}
\vglue 0.1cm

\rightskip=0pc
\leftskip=0pc
\vskip 0.5cm
{\centerline{\elevenit 2.6 Model independent analyses
and model building constraints}}
\vskip 0.5cm

In this lecture we have revised the parametrization of an
extra $U(1)$ interaction, specifying the values of the
new couplings for typical models; and we have introduced
convenient sets of normalized charges for comparison with
experiment at hadron and lepton colliders. To finish we
discuss some predictions for these couplings from
unification near the Planck scale.

Models with an extra abelian gauge interaction are fixed
by two extra gauge couplings: $g_2, g_{12}$. Their values
near $M_Z$ result from their evolution from the Planck
down to the electroweak scale. Hence they are a window
to all intermediate scales and matter contents in between.
For illustration we gathered in Table 5 the
$\lambda , \theta _1$ values at the electroweak scale
for some (grand) unified models [12], as well as
the corresponding $g_2, g_{12}$ values.
In all cases $g_{12}=0$ at the unification scale.
We also quote the values for the $\chi $ and $LR$ models.
The (grand) unified models assume a minimal matter content
and $\alpha _C=0.12, \alpha _E=128^{-1}$ and $s_W^2=0.23$
(which fixes the intermediate scale) at $M_Z$. The required
intermediate and grand unified scales for these examples
are probably too
low/large, but there are many non-minimal matter contents
which give acceptable models and a large variety of
$\lambda , \theta _1\ (g_2, g_{12})$ values [13]. The
$\chi , LR$ models with a minimal matter content and no
intermediate scale do not unify. In conclusion, if a new
$Z'$ exists and then $\lambda , \theta _1$ can be measured,
they will provide precious information on the physics at
and well beyond the $TeV$ scale.

\vglue 0.2cm
\begin{center}
\begin{tabular}{|c|cccc|}
\hline
\multicolumn{1}{|c}{$G\rightarrow SU(3)_C\times
SU(2)_L\times U(1)_Y\times U(1)_{\chi}$} & $\lambda$
& $\theta _1$ & $\frac {g_2}{\sqrt {\frac {5}{3}}
\frac {e}{c_W}}$ &
\multicolumn{1}{c|}{$\frac {g_{12}}{g_2}$} \\
\hline
$ B3: SU(4)_C\times
SU(2)_L\times SU(2)\times U(1) $ & $ 1.37 $ & $ 1.20 $ &
$ 1.28 $ & $ 0.17 $ \\
$ B4: SU(3)_C\times
SU(3)_L\times SU(2)\times U(1) $ & $ 1.18 $ & $ -1.02 $ &
$ 1.09 $ & $ -0.14 $ \\
$ B7: SU(4)_C\times
SU(2)_L\times SU(2)_R\times U(1) $ & $ 1.08 $ & $ 0.17 $ &
$ 0.93 $ & $ 0.03 $ \\
$ B8: SU(3)_C\times
SU(3)_L\times SU(2)_R\times U(1) $ & $ 1.23 $ & $ -0.69 $ &
$ 0.98 $ & $ -0.21 $ \\
$ C1, ..., C5, D: SU(3)_C\times
SU(2)_L\times U(1)_Y\times U(1)_{\chi} $
& $ 1.04 $ & $ -0.89 $ &
$ 1.01 $ & $ -0.04 $ \\
 & & & & \\
$ \chi $ & $ 1 $ & $ - $ &
$ 1 $ & $ 0 $ \\
$ LR $ & $ 1.88 $ & $ 0.68 $ &
$ 1.07 $ & $ 0.62 $ \\
\hline
\end{tabular}
\end{center}
\vglue 0.1cm

\vglue 0.1cm
\rightskip=3pc
\leftskip=3pc
{\tenrm \baselineskip=12pt
\noindent
Table 5. $\lambda , \theta _1$ values and the corresponding
gauge couplings $g_2, g_{12}$ for some unified models and
for the $\chi , LR$ models.}
\vglue 0.1cm

\rightskip=0pc
\leftskip=0pc
\vglue 1.cm
{\centerline{\elevenbf 3. Indirect limits on extra gauge
interactions}}
\vglue 0.5cm

We give a short summary of the second lecture.
At present there are only limits on new weak interactions.
These are direct bounds on $Z'$ production at TEVATRON
(see next lecture) and indirect bounds from fits to
precise electroweak data. (We do not consider cosmological
bounds in these lectures.) Indirect limits are very model
dependent. The experimental precision is at the level of
radiative corrections and new effects must be very small.
Hence, one must wonder not only about modifications of tree
level expressions but about extra loop contributions.
Enlarging the standard model consistently often requires
to add many new particles and then many new parameters.
As a matter of fact often indirect $Z'$ limits must be
understood as an order of magnitude estimate.

In practice the extra interaction is taken into account
in fits at tree level only. This approximation is
better when the limits are more stringent [14]. There are
two approaches for reducing the number of free parameters:
stick with simple, well-defined models and/or use
phenomenologically relevant parameters. An example of the
first approach can be found in Ref. [15]. The simplest
class of extended models within $E_6$ is the class
defined by the extra charge within $SO(10): Q_{\chi}=
Q_{LR}$ in Table 3. These models contain right-handed
neutrinos, which we assume to be heavy, and a minimal
Higgs sector which makes the $Z_1Z'$ mixing a well-defined
function of the $Z'$ mass. The class is then fixed by
the $Z'$ mass, $M_{Z'}$, and two gauge couplings, $g_2,
g_{12}$ parametrized by $\lambda ,  \theta _1$ (see Eq.
(5)). See Table 5 for examples.
In Ref. [15] the $M_{Z'}$ lower bounds
($90 \% c.l.$ for one variable) implied by 1991 precise
electroweak data are plotted as a function of $\lambda $
and $\theta _1$.
The standard model limit corresponds to $\theta _1=0,
\lambda \rightarrow \infty $.

The second approach prefers a parametrization adapted
to available data. Low energy electroweak data as well
as LEP data are mainly sensitive to the $Z'$ mass and the
$Z_1Z'$ mixing. These redefine the $\rho $ parameter and
the $Z$ current:

\begin{equation}
\begin{array}{l}
{\displaystyle
\rho = (1-\frac {s_3^2(M^2_{Z'}-M^2_Z)c_W^2}{M_W^2})^{-1},}
\\
{\displaystyle
J_Z=c_3J_{Z_1}-s_3J_{Z_2}.}
\end{array}
\end{equation}

\noindent
$\rho $ ($J_Z$) is mainly constrained by low energy (LEP)
data. A detailed analysis is more involved. The present limits
can be summarized saying that once fixed the $Z'$ charges,
for a large class of models present data require [9,16]

\begin{equation}
M_{Z'} > 200\ GeV,\ s_3 < 0.01.
\end{equation}

\noindent
See T. Riemann lectures for a detailed discussion of these
bounds, in particular from LEP [2].

\vglue 1.cm
{\centerline{\elevenbf 4. Determination of ${\bf Z'}$
gauge couplings to ordinary fermions at future colliders}}
\vglue 0.5cm

In this lecture we study the potential of large hadron and
lepton colliders for the discovery of a new neutral gauge
boson and for a model independent determination of its
couplings. We use popular models for illustration only.

\vskip 0.5cm
{\centerline{\elevenit 4.1 Hadron colliders}}
\vskip 0.5cm

If heavy gauge bosons turn out to have a mass in the few
$TeV$ region, future hadron colliders, $e.g.$ LHC, would
be an ideal place to discover and study them. In the main
production channels, $p^(\bar p^)\rightarrow Z'\rightarrow
l\bar l (l=e,\mu)$, one would be able to measure

\begin{itemize}
\item the mass $M_{Z'}$,
\item the width $\Gamma _{Z'}$ and
\item the total cross section $\sigma _{ll}$, as well as
\item the forward-backward asymmetry $A_{FB}$ [5,17] and
\item the ratio of cross-sections in different rapidity
bins $r_{y_1}$ [10,18].
\end{itemize}

\noindent
The $Z'$ can also decay into $WW$ and $ZH$. These
decays are observed, however, as four fermion final
states for $W, Z, H$ decay into two fermions. (A heavy
Higgs decays mainly into two gauge bosons.) Hence,
the next dominant $Z'$ decays are those involving four
fermions. They have small cross sections and are difficult
to isolate in general, but they are important to determine
the $Z'$ couplings [10]. The main probes are

\begin{itemize}
\item rare decays $W\rightarrow Wl\nu _l$ [19,20] and
\item associated productions $pp\rightarrow Z'V$ with
$V=Z, W, \gamma $ [21].
\end{itemize}

\vskip 0.5cm
{\centerline{\elevenit 4.1.1 Two fermion final states: $p^(\bar p^)
\rightarrow \gamma , Z, Z'\rightarrow l\bar lX;
jet jet X$}}
\vskip 0.5cm

The unpolarized differential cross section for the process
$p^(\bar p^)\rightarrow
\gamma , Z, Z'\rightarrow l\bar lX$ (with $l$ a definite
charged lepton)
depends on the lepton-antilepton invariant mass $M$,
where $M^2=(p_l+p_{\bar l})^2$,
on the rapidity $y$, where $\frac {M}{2}(e^y-e^{-y})=
(p_l+p_{\bar l})_{longitudinal}$, and on $\theta ^*$,
which is the scattering angle $pl$ in
the center of mass of the parton system $q\bar q$ [5,17].
The general form of
this cross section is, for tree level amplitudes,
(quark and lepton masses are neglected)

\begin{equation}
\frac {d\sigma }{dMdydcos\theta ^*}=
{\sum _{q=u,c,t,d,s,b}}\frac {M}{192\pi}
[g^S_q(y,M)S_q(M)(1+cos^2\theta ^*)+
g^A_q(y,M)A_q(M)2cos\theta ^*],
\end{equation}

\noindent
where $S_q, A_q$ (which are the only model dependent
quantities) involve the gauge couplings to
quarks and leptons,

\begin{equation}
\begin{array}{l}
{\displaystyle
S_q ={\sum _{\alpha ,\beta =\gamma ,Z,Z'}}
{\frac {(g_{L\alpha}^qg_{L\beta}^q+
g_{R\alpha}^qg_{R\beta}^q)(g_{L\alpha}^lg_{L\beta}^l+
g_{R\alpha}^lg_{R\beta}^l)}
{(M^2-M^2_{\alpha}+iM_{\alpha}\Gamma _{\alpha})
(M^2-M^2_{\beta}-iM_{\beta}\Gamma _{\beta})}}},
\\
{\displaystyle
A_q ={\sum _{\alpha ,\beta =\gamma ,Z,Z'}}
{\frac {(g_{L\alpha}^qg_{L\beta}^q-
g_{R\alpha}^qg_{R\beta}^q)(g_{L\alpha}^lg_{L\beta}^l-
g_{R\alpha}^lg_{R\beta}^l)}
{(M^2-M^2_{\alpha}+iM_{\alpha}\Gamma _{\alpha})
(M^2-M^2_{\beta}-iM_{\beta}\Gamma _{\beta})}}},
\end{array}
\end{equation}

\noindent
and $g_q^S, g_q^A$ are the parton distribution
functions of the colliding hadrons [22],

\begin{equation}
g_q^{S,A}(y,M)=x_ax_b[f^{(a)}_q(x_a,M^2)
f^{(b)}_{\bar q}(x_b,M^2)\pm f^{(a)}_{\bar q}(x_a,M^2)
f^{(b)}_q(x_b,M^2)],
\end{equation}

\noindent
where the sign $\pm $ correspond to $S,A$.
$a$ and $b$ are the two colliding hadrons at the center
of mass energy $\sqrt s$; $x_a$ and $x_b$ being the
momentum fractions of the colliding partons in $a$ and
$b$ respectively, $x_{a,b}=\frac {M}{\sqrt s}e^{\pm y}$.
The gauge couplings were discussed in the first lecture:

\begin{equation}
\begin{array}{l}
{\displaystyle
g^i_{L,R\ \gamma}\equiv eq_i},
\\
{\displaystyle
g^i_{L,R\ Z}\equiv \frac {g}{2c_W}[(v^ic_3-v'^is_3)\pm
(a^ic_3-a'^is_3)]},
\\
{\displaystyle
g^i_{L,R\ Z'}\equiv g_2\hat g^i_{L,R\ Z'}
=\frac {g}{2c_W}[(v'^ic_3+v^is_3)\pm
(a'^ic_3+a^is_3)]}.
\end{array}
\end{equation}

\noindent
At large hadron colliders the $\gamma , Z$ contributions
and the $Z_1Z'$ mixing can be neglected in two fermion
channels. The statistics
eventually available and the limit on $s_3$ from precise
electroweak data discussed in the previous lecture allow
for considering only the $Z'$ contribution (no
$\gamma , Z$ interference) with zero gauge boson mixing
($s_3=0$). Then Eqs. (9, 10) apply.

In the absence of a positive
signal the $Z'$ total cross section

\begin{equation}
\sigma _{ll}=\int ^{\sqrt s}_0 dM \int
^{ln\frac {\sqrt s}{M}}_{ln\frac {M}{\sqrt s}}
dy \int ^1_{-1}dcos\theta ^*
\frac {d\sigma }{dMdydcos\theta ^*}, \
({\rm with}\ \alpha ,\beta =Z'\ {\rm in\ Eq.\ (15)},)
\end{equation}

\noindent
for each given model (with the gauge coupling strength and
$\Gamma _{Z'}$ also fixed) puts a limit on $M_{Z'}$.
TEVATRON limits on an excess of isolated lepton pairs with
large invariant mass translate into limits on $M_{Z'}$.
In Table 6 we gather the $Z'$ couplings to ordinary
fermions and the minimal $Z'$ width for the popular models
discussed in the first lecture.
For an integrated luminosity of $4\ pb^{-1}$ the
$M_{Z'}$ ($95\% \ c.l.$) bounds for these models are
[6,23,24]

\begin{equation}
M_{\chi , \psi , \eta , LR} > 340, 320, 340, 310\ GeV.
\end{equation}

\vglue 0.2cm
\begin{center}
\begin{tabular}{|c|cccc|}
\hline
\multicolumn{1}{|c}{} & $\chi$
& $\psi$ & $\eta$ &
\multicolumn{1}{c|}{$LR$} \\
\hline
$g_2$
& $ 0.461 $ & $ 0.461 $ & $ 0.461 $ & $ 0.493 $ \\
$ $ & $ $ & $ $ & $ $ & $ $ \\
$\hat g^l_{L\ Z'}=\hat g^{\nu}_{L\ Z'}=\hat g^e_{L\ Z'}$
& $ \frac {3}{2\sqrt {10}} $ &
$ \frac {1}{2\sqrt {6}} $ & $ \frac {1}{2\sqrt {15}} $ &
$ 0.253 $ \\
$\hat g^e_{R\ Z'}$ & $ \frac {1}{2\sqrt {10}} $
& $ -\frac {1}{2\sqrt {6}} $ & $ \frac {1}{\sqrt {15}} $
& $ -0.341 $ \\
$\hat g^q_{L\ Z'}=\hat g^u_{L\ Z'}=\hat g^d_{L\ Z'}$
& $ -\frac {1}{2\sqrt {10}} $
& $ \frac {1}{2\sqrt {6}} $ & $ -\frac {1}{\sqrt {15}} $
& $ -0.084 $ \\
$\hat g^u_{R\ Z'}$ & $ \frac {1}{2\sqrt {10}} $
& $ -\frac {1}{2\sqrt {6}} $ & $ \frac {1}{\sqrt {15}} $
& $ 0.509 $ \\
$\hat g^d_{R\ Z'}$ & $ -\frac {3}{2\sqrt {10}} $
& $ -\frac {1}{2\sqrt {6}} $ & $ -\frac {1}{2\sqrt {15}} $
& $ -0.678 $ \\
$ $ & $ $ & $ $ & $ $ & $ $ \\
$\frac {\Gamma _{Z'}}{M_{Z'}}$
& $0.012$ & $0.006$ & $0.007$ & $0.021$ \\
\hline
\end{tabular}
\end{center}
\vglue 0.1cm

\vglue 0.1cm
\rightskip=3pc
\leftskip=3pc
{\tenrm \baselineskip=12pt
\noindent
Table 6. Popular $Z'$ model couplings
$g^i_{L,R\ Z'}=g_2\hat g^i_{L,R\ Z'}$ and widths
$\Gamma _{Z'}$.}
\vglue 0.1cm

\rightskip=0pc
\leftskip=0pc
If a new gauge boson with a mass $\leq 5\ TeV$ exits
and couples to ordinary fermions with a sizeable
strength ($g_2 \sim 0.1$), it
should be observed in the two lepton channel at LHC.
Once a new $Z'$ is observed
in the two lepton channel
and its mass $M_{Z'}$ and width $\Gamma _{Z'}$
measured, one must try to measure the $Z'$ couplings to
fermions in a model independent way. With no $\gamma , Z$
interference no determination of the signs of the $Z'$
charges is possible (see Eqs. (14,15)). Therefore
it is convenient to use the gauge coupling strength $g_2$
and four normalized couplings with no sign ambiguity, $e.g.
\ \gamma _L^l, \gamma _L^q, \tilde U, \tilde D$ in Eq. (10)
(their values for typical models were given in Table 4),
as free parameters. What matters is which couplings can be
measured at LHC. In the two lepton channel the relevant
observables are the total cross section
$\sigma _{ll}$, the forward-backward
asymmetry

\begin{equation}
\begin{array}{cc}
{\displaystyle
A _{FB}} &
{\displaystyle
=\frac {\int dM (\int ^{ln\frac {\sqrt s}{M}}_0
- \int ^0_{ln\frac {M}{\sqrt s}})
dy (\int ^1_0 - \int ^0_{-1})dcos\theta ^*
\frac {d\sigma }{dMdydcos\theta ^*}}
{\sigma _{ll}}}
\\
& {\displaystyle
=0.38(2\gamma ^l_L-1)
\frac {1-0.75\tilde U-0.25\tilde D}
{1+0.68\tilde U+0.32\tilde D}},
\end{array}
\end{equation}

\noindent
and the ratio of cross-sections in different rapidity
bins ($y_1$ is chosen in a range $0 < y_1 < y_{max}$
so that the number of events in the two bins are
comparable, $y_1=1$ at LHC where $\sqrt s=16\ TeV$)
[10,18]

\begin{equation}
\begin{array}{cc}
{\displaystyle
r _{y_1}}
&{\displaystyle
=\frac {\int dM
\int _{-y_1}^{y_1} dy \int dcos\theta ^*
\frac {d\sigma }{dMdydcos\theta ^*}}
{\int dM (\int _{-y_{max}}^{-y_1} +
\int _{y_1}^{y_{max}}) dy \int dcos\theta ^*
\frac {d\sigma }{dMdydcos\theta ^*}}}
\\
&{\displaystyle
=1.55
\frac {1+0.64\tilde U+0.36\tilde D}
{1+0.73\tilde U+0.27\tilde D}}.
\end{array}
\end{equation}

\noindent
Another observable is the ratio

\begin{equation}
\begin{array}{cc}
{\displaystyle
A_{FB_{y_1}}}
&{\displaystyle
=\frac {\int dM
(\int _0^{y_1}-\int _{-y_1}^0) dy (\int ^1_0 -
\int ^0_{-1}) dcos\theta ^*
\frac {d\sigma }{dMdydcos\theta ^*}}
{\int dM (\int _{y_1}^{y_{max}} -
\int _{-y_{max}}^{-y_1}) dy (\int ^1_0 -
\int ^0_{-1}) dcos\theta ^*
\frac {d\sigma }{dMdydcos\theta ^*}}}
\\
&{\displaystyle
=0.60
\frac {1-0.73\tilde U-0.27\tilde D}
{1-0.76\tilde U-0.24\tilde D}}.
\end{array}
\end{equation}

\noindent
The numerical expressions above are calculated assuming
$M_{Z'}=1\ TeV$ and the EHLQ structure functions set 1
[22]. A numerical analysis shows that $A_{FB_{y_1}}$ is
not a sensitive enough function of the gauge couplings to
provide useful information for the projected
luminosities [10].
Thus only three combinations out of five gauge couplings
can be determined in the two lepton channel. In particular
$A _{FB}, r _{y_1}$, and $A_{FB_{y_1}}$ fix two combinations
out of the four normalized
couplings. Other observables do not provide more
information if the $Z'$ interference with the $\gamma , Z$
bosons is neglected [18]. At any rate, a poor energy
resolution would erase any interference effects from
data.

The two lepton final state does not constrain
$\gamma ^q_L$. This process is proportional to the product
of the $Z'$ coupling to quarks times the $Z'$ coupling to
leptons. Hence, a simultaneous increase (decrease) of the
quark coupling and a decrease (increase) of the lepton
coupling by the same factor does not change the $Z'$ cross
section into leptons at LHC. The relative size of the $Z'$
couplings to quarks and leptons could be determined by a
measurement of the $Z'$ cross section into quark pairs. In
particular, the ratio

\begin{equation}
\frac {1}{3}\frac {\sigma (pp\rightarrow Z' \rightarrow q
\bar q)}{\sigma (pp\rightarrow Z' \rightarrow l^+l^-)} =
\gamma ^q_L (2+\tilde U+\tilde D)
\end{equation}

\noindent
(counting all three families) would yield the left-handed
quark coupling $\gamma ^q_L$ [10]. However, this appears
difficult [25].

The $\tau$ polarization in $pp\rightarrow Z' \rightarrow
\tau ^+\tau ^-$ would be another useful probe if it can be
measured [26].

Similarly, if proton polarization were available the
measurements of the corresponding asymmetries in
$pp\rightarrow Z' \rightarrow l^+l^-$ would also be
useful [27].

\vskip 0.5cm
{\centerline{\elevenit 4.1.2 Four fermion final states: $Z'
\rightarrow WW, ZH, f\bar fV; Z'V$}}
\vskip 0.5cm

A new $Z'$ with a mass $\leq 2\ TeV$ can also show up in four
fermion decays [19,21,28]. The $WW, ZH$ decays have a rate
similar to the $Z'$ decay into two fermions plus one heavy
gauge boson, $f\bar f V, V=W,Z$. All of them are observed
as four fermion decays; and then they must be considered
together because they can interfere. Associated production
$Z'V, V=Z,W,\gamma ,$ cross sections are of a similar size,
too.

The $Z'$ four fermion cross sections are relatively small
on one hand, and on the other it is difficult to isolate
the corresponding samples. More detailed analyses
(simulations) are still needed to decide on the actual
relevance of these processes. Below we discuss the main
four fermion cross sections, assuming full
efficiency in the identification of the final states,
and perform a fit to know how well the $Z'$ couplings
could be ideally measured, including only statistical
errors. After, we comment on the isolation of four
fermion signals.

\begin{itemize}
\item $Z'\rightarrow WW$ [28]. The $Z'$ decay rate into $W$
pairs is proportional to the  square of the trilinear
$Z'WW$ coupling $\delta _{Z'WW}$:

\begin{equation}
\Gamma (Z'\rightarrow WW) = \frac {M_{Z'}}{12\pi}
\frac {\delta ^2_{Z'WW}}{16\eta ^2_W}
(1-4\eta _W)^{\frac {3}{2}}(1+20\eta _W+12\eta _W^2),
\end{equation}

\noindent
with $\eta _W\equiv \frac {M_W^2}{M_{Z'}^2}$. This
trilinear coupling is equal to the $Z_1Z'$ mixing
times the $Z_1WW$ coupling, which is the product of
the $SU(2)_L$ coupling times the $W_3Z_1$ mixing, [29]:

\begin{equation}
\delta _{Z'WW} = s_3\frac {e}{s_W}c_W.
\end{equation}

\noindent
In typical models $s_3\sim \frac {M_Z^2}{M_{Z'}^2}$
and $\Gamma (Z'\rightarrow WW)$ is not very large.
$pp\rightarrow Z' \rightarrow WW$ is the only process
probing directly the gauge symmetry breaking coupling
$\delta _{Z'WW}$ and then $s_3$. As $W$'s are observed
through their decays, to isolate this signal requires
refined cuts (see below).
\item $Z'\rightarrow ZH$ [30,31]. This decay is somewhat
model dependent. We will not consider it here.
\item $Z'\rightarrow f\bar fV$ [19]. Recently it has been
emphasized that this decay rate is enhanced due to the
collinear singularity associated to soft vector boson
emission:

\begin{equation}
\Gamma (Z'\rightarrow f\bar fZ) = \frac
{M_{Z'}(g_{LZ}^{f\ 2}g_{LZ'}^{f\ 2}+
g_{RZ}^{f\ 2}g_{RZ'}^{f\ 2})}{192\pi ^2}[ln^2\eta _Z +
3ln\eta _Z + 5 - \frac {\pi ^2}{3} + O(\eta _Z)],
\end{equation}

\noindent
with $\eta _Z\equiv \frac {M_Z^2}{M_{Z'}^2} << 1$.
$\Gamma (Z'\rightarrow f\bar fZ)$ has a similar
expression but $M_Z$ must be replaced by $M_W$ and the
$Z$ couplings by $g_{LW}^f=\frac {e}{\sqrt 2s_W},
g_{RW}^f=0$. The photon emission decay $Z'\rightarrow
f\bar f \gamma $ is a correction of the $Z'$ decay
into $f\bar f$ and it does not provide new information
on the $Z'$ coupling to fermions.
\item {\elevenit Associated $Z'$ production: $Z'V$} [21].
These cross sections are typically $1\%$ the $Z'$ production
cross section. The different electric charge of up
and down quarks makes $Z'\gamma $ production at LHC a
useful probe to measure the $Z'$ couplings to quarks.
\end{itemize}

The only rare decay which provides new information on
$Z'$ couplings to fermions and which seems possible to
isolate is $Z'\rightarrow l\nu W$ (see, however, Ref.
[32]). Normalizing to the $Z'$ decay into lepton pairs
($l=e, \mu $) and assuming $M_{Z'}=1\ TeV$ and the EHLQ
structure functions set 1, the four fermion probes
can be written (as a function of the normalized charges
defined above) at LHC [10]:

\begin{equation}
\begin{array}{c}
{\displaystyle
r _{l\nu W}
\equiv \frac {\sigma (pp\rightarrow Z'\rightarrow
(W^{\pm}\rightarrow hadrons)l^{\mp}\nu )}
{\sigma (pp\rightarrow Z'\rightarrow l^+l^-)}}
\\
{\displaystyle
=0.067\gamma ^l_L},\\
{\displaystyle
R _{Z'Z}
\equiv \frac {\sigma (pp\rightarrow (Z'\rightarrow
l^+l^-)Z)}
{\sigma (pp\rightarrow Z'\rightarrow l^+l^-)}}
\\
{\displaystyle
=10^{-3}\frac {7.94+0.96\tilde U+0.11\tilde D}
{1+0.68\tilde U+0.32\tilde D}},\\
{\displaystyle
R _{Z'W}
\equiv \frac {\sigma (pp\rightarrow (Z'\rightarrow
l^+l^-)W)}
{\sigma (pp\rightarrow Z'\rightarrow l^+l^-)}}
\\
{\displaystyle
=10^{-3}\frac {25.7}
{1+0.68\tilde U+0.32\tilde D}},\\
{\displaystyle
R _{Z'\gamma }
\equiv \frac {\sigma (pp\rightarrow (Z'\rightarrow
l^+l^-)\gamma )}
{\sigma (pp\rightarrow Z'\rightarrow l^+l^-)}}
\\
{\displaystyle
=10^{-3}5.62\frac {1+0.89\tilde U+0.11\tilde D}
{1+0.68\tilde U+0.32\tilde D}}
\end{array}
\end{equation}

\noindent
(with a transverse momentum cut $p_t^{\gamma } > 50\ GeV$).
As can be readily seen these expressions do not contain
(constrain) $\gamma ^q_L$, because they are quadratic
in the $Z'$ couplings to quarks as well as to leptons.
Hence when fitted with the two lepton $Z'$ observables,
they fix three out of four normalized couplings, leaving
$\gamma ^q_L$ unconstrained. In Table 7 we write the
values of these observables for typical models. Errors
are statistical only. (We assume an integrated luminosity
of $10^5 pb^{-1}$.) Performing a fit to these values
we can determine $\gamma ^l_L, \tilde U, \tilde D$. The
results are given in Table 8.

\vglue 0.2cm
\begin{center}
\begin{tabular}{|c|cccc|}
\hline
\multicolumn{1}{|c}{} & $\chi$
& $\psi$ & $\eta$ &
\multicolumn{1}{c|}{$LR$} \\
\hline
$A_{FB}$ & $ -0.134\pm 0.007 $ & $ 0.0\pm 0.016$ &
$ -0.025\pm 0.014 $ & $ 0.098\pm 0.006 $ \\
$r_{y_1}$ & $ 1.79\pm 0.02 $ & $ 1.55\pm 0.04 $ &
$ 1.49\pm 0.03 $ & $ 1.62\pm 0.014 $ \\
$A_{FB_{y_1}}$ & $ 0.68\pm 0.08 $ & $ $ &
$ 0.68\pm 0.88 $ & $ 0.61\pm 0.08 $ \\
 & & & & \\
$r_{l\nu W}$ & $ 0.060\pm 0.0014 $ & $ 0.034\pm 0.002 $ &
$ 0.013\pm 0.001 $ & $ 0.024\pm 0.0008 $ \\
$R_{Z'Z}$ & $ 0.0022\pm 0.0002 $ & $ 0.0045\pm 0.0008 $ &
$ 0.0051\pm 0.0007 $ & $ 0.0011\pm 0.0001 $ \\
$R_{Z'W}$ & $ 0.0056\pm 0.0004 $ & $ 0.013\pm 0.001 $ &
$ 0.015\pm 0.001 $ & $ 0.00055\pm 0.00010 $ \\
$R_{Z'\gamma}$ & $ 0.0035\pm 0.0003 $ & $ 0.0056\pm
0.0009$ & $ 0.0061\pm 0.0008 $ & $ 0.0049\pm 0.0003 $ \\
\hline
\end{tabular}
\end{center}
\vglue 0.1cm

\vglue 0.1cm
\rightskip=3pc
\leftskip=3pc
{\tenrm \baselineskip=12pt
\noindent
Table 7. Values of two and four fermion $Z'$ observables
for popular models. Errors are statistical only. Error
bars for $r_{y_1}, r_{l\nu W}, R_{Z'V}$ are for $e$ plus
$\mu$ channels, while $A_{FB}, A_{FB_{y_1}}$ are for $e$
or $\mu$.}
\vglue 0.1cm

\rightskip=0pc
\leftskip=0pc
\vglue 0.2cm
\begin{center}
\begin{tabular}{|c|cccc|}
\hline
\multicolumn{1}{|c}{} & $\chi$
& $\psi$ & $\eta$ &
\multicolumn{1}{c|}{$LR$} \\
\hline
$\gamma ^l_L$ & $ 0.9\pm 0.018 $ & $ 0.5\pm 0.03 $ &
$ 0.2\pm 0.015 $ & $ 0.36\pm 0.007 $ \\
$\gamma ^q_L$ & $ 0.1 $ & $ 0.5 $ &
$ 0.8 $ & $ 0.04 $ \\
$\tilde U$ & $ 1\pm 0.18 $ & $ 1\pm 0.27 $ &
$ 1\pm 0.14 $ & $ 37\pm 8.3 $ \\
$\tilde D$ & $ 9\pm 0.61 $ & $ 1\pm 0.41 $ &
$ 0.25\pm 0.29 $ & $ 65\pm 14 $ \\
\hline
\end{tabular}
\end{center}
\vglue 0.1cm

\vglue 0.1cm
\rightskip=3pc
\leftskip=3pc
{\tenrm \baselineskip=12pt
\noindent
Table 8. Values of $\gamma ^l_L, \gamma ^q_L, \tilde U,
\tilde D$ for the $\chi , \psi , \eta , LR$ models. The
error bars indicate how well the coupling could be
measured at the LHC for $M_{Z'}=1\ TeV$.}
\vglue 0.1cm

\rightskip=0pc
\leftskip=0pc
Four fermion $Z'$ decays have large backgrounds. No
detailed simulation of the different $Z'$ rare decays
has been published. To illustrate the reduction factors
we have to pay to isolate a relatively clean sample,
let us discuss the $Z'$ decay into an electron and a
muon plus missing energy [20]. To this signal contribute
$Z'\rightarrow (W\rightarrow e\nu )(W\rightarrow \mu \nu ),
(W\rightarrow e\nu )\mu \nu , (W\rightarrow \mu \nu )e\nu $.
Wondersome backgrounds are $W^+W^-$ continuum, $Z'
\rightarrow \tau \bar {\tau }$, and (QCD) $t\bar t$
production. In Fig. 1 of Ref. [20] there are plotted
the missing transverse
momentum, ${\not p}_t$, and $e\mu $ angle in the transverse
plane, $\theta ^{e\mu }_t$, distributions. They are
simulations at the parton level. The signal corresponds
to the $\chi $ model with $M_{Z'}=1\ TeV$. The top
background was estimated assuming a top with a
mass of $150\ GeV$ and that $t$ decays into $bW$,
and $W$ into $l\nu $, isotropically. How large the top
quark background might be depends on how good the top
quark reconstruction is. Lepton isolation is enforced
demanding the distance in the lego plot

\begin{equation}
\Delta R_{ql}={\sqrt {(\Delta \phi )^2 + (\Delta \eta )
^2}} > 0.75;
\end{equation}

\noindent
whereas misidentification of the top (bottom) quark decay
is taken into account requiring $p_t^{b,\bar b} < 50\ GeV$
and a good balancing of the transverse momentum of the
sample

\begin{equation}
\vec p_t^{\ out} \equiv (\vec p_e+\vec p_{\mu }+
{\not {\vec p}}) _t < 25\ GeV.
\end{equation}

\noindent
The ${\not p} _t$ and $\theta _t^{e\mu }$ distributions
examplify what we want to emphasize: $Z'$ rare
decays have large backgrounds at LHC. In particular,
the extremely large QCD cross sections require, in order
to isolate the signals, identification of isolated
leptonic tracks and stringent cuts on transverse
energy. To obtain a relatively clean $Z'\rightarrow
e\mu {\not p}$ sample a set of cuts, which reduces
the signal by at least a factor of 3, is needed; $e.g.$
the 450 events produced reduce to 120 after requiring
${\not p} _t > 200\ GeV, p_t^{e, \mu } > 50\ GeV,
cos \theta _t^{e\mu } > -0.95$.

\vskip 0.5cm
{\centerline{\elevenit 4.2 Lepton colliders}}
\vskip 0.5cm

If a new $Z'$ exists with a mass much larger than the
available center of mass energy, its main effects
at a large lepton collider result from its interference
with $\gamma , Z$. (The (tree level)
cross section for $e^+e^-\rightarrow \gamma , Z, Z'
\rightarrow f\bar f$ can be found in Ref. [33].) This
interference allows for measuring all $Z'$ gauge
couplings to ordinary fermions, including their
relative signs. A convenient set of four normalized
charges, $P_V^l, P_L^q, P_R^u, P_R^d$, was defined
in Eq. (11). A fit to $e^+e^-\rightarrow
f\bar f, f=l, u, d,$ observables is presented for
$M_{Z'} = 1\ TeV$ at NLC (center of mass
energy $\sqrt s = 500\ GeV$ and integrated luminosity
${\cal L}_{int} = 20 fb^{-1}$)
in Ref. [11]. (See also Ref. [34].) The values used in the
fit for the different observables correspond to typical
models. Errors are only statistical. The results of the fit
are gathered in Table 9.
$100\%$ efficiency for the heavy flavour tagging and
$100\%$ polarization for the electron beam are assumed.
(Error bars in parentheses correspond to unpolarized
electron beams.)
\eject

\vglue 0.2cm
\begin{center}
\begin{tabular}{|c|cccc|}
\hline
\multicolumn{1}{|c}{} & $\chi$
& $\psi$ & $\eta$ &
\multicolumn{1}{c|}{$LR$} \\
\hline
$ P_V^e $ & $ 2\pm 0.08(0.26) $ & $ 0\pm 0.04(1.5) $ &
$ -3\pm 0.5(1.1) $ & $ -0.15\pm 0.018(0.072) $ \\
$ P_L^q $ & $ -0.5\pm 0.04(0.10) $ & $ 0.5\pm 0.10(0.2) $ &
$ 2\pm 0.3(1.1) $ & $ -0.14\pm 0.037(0.07) $ \\
$ P_R^u $ & $ -1\pm 0.15(0.19) $ & $ -1\pm 0.11(1.2) $ &
$ -1\pm 0.15(0.24) $ & $ -6.0\pm 1.4(3.3) $ \\
$ P_R^d $ & $ 3\pm 0.24(0.51) $ & $ -1\pm 0.21(2.8) $ &
$ 0.5\pm 0.09(0.48) $ & $ 8.0\pm 1.9(4.1) $ \\
\hline
\end{tabular}
\end{center}
\vglue 0.1cm

\vglue 0.1cm
\rightskip=3pc
\leftskip=3pc
{\tenrm \baselineskip=12pt
\noindent
Table 9. Values of the couplings $P_V^e, P_L^q, P_R^u,
P_R^d$ and errors as determined at NLC. $100\%$ heavy
flavour tagging efficiency and $100\%$ longitudinal
polarization of the electron beam are assumed for the
first set of error-bars, while the error-bars in
parentheses are for the probes without polarization.}
\vglue 0.1cm

\rightskip=0pc
\leftskip=0pc
\vglue 1.cm
{\centerline{\elevenbf 5. Concluding remarks}}
\vglue 0.5cm

In these lectures we have presented a general
parametrization of an extra abelian, gauge interaction,
we have commented on present limits on new gauge
bosons, and we have discussed the $Z'$ potential
of LHC and NLC. Both colliders are complementary.
(We did not consider $t$-channel limits from HERA.
They are not competitive with the $s$-channel limits
from large hadron and lepton colliders.) If a new
$Z'$ exists in the $TeV$ range, it should be observed
at LHC, where its mass and width can be measured. At a
large hadron collider the strength of the new interaction
and three out of four normalized charges fixing the
$Z'$ couplings to ordinary fermions can be determined
for $M_{Z'} \leq 2\ TeV$. The determination of their
relative signs is difficult. At NLC all the $Z'$ couplings
to quarks and leptons can be measured, including their
relative signs. The average errors tend to be larger
than at LHC,
however. In Table 10 we give the errors for the normalized
observables defined at LHC, which are implied by the
NLC analysis in Table 9. These errors have to be compared
with the errors expected at LHC in Table 8.
For other reviews on $Z'$ physics see Refs. [31,35].

\vglue 0.2cm
\begin{center}
\begin{tabular}{|c|cccc|}
\hline
\multicolumn{1}{|c}{} & $\chi$
& $\psi$ & $\eta$ &
\multicolumn{1}{c|}{$LR$} \\
\hline
$\gamma ^l_L$ & $ 0.9\pm 0.010(0.031) $ & $ 0.5\pm
0.04(1.5) $ & $ 0.2\pm 0.04(0.09) $ & $ 0.36\pm
0.02(0.07)$ \\
$\gamma ^q_L$ & $ 0.1\pm 0.017(0.045) $ & $ 0.5\pm
0.20(0.4)$ & $ 0.8\pm 0.34(1.03) $ & $ 0.04\pm
0.02(0.04)$ \\
$\tilde U$ & $ 1\pm 0.30(0.38) $ & $ 1\pm 0.22(2.4) $ &
$ 1\pm 0.30(0.48) $ & $ 37\pm 16.8(39.6) $ \\
$\tilde D$ & $ 9\pm 1.44(3.06) $ & $ 1\pm 0.42(5.6) $ &
$ 0.25\pm 0.09(0.48) $ & $ 65\pm 30.4(65.6) $ \\
\hline
\end{tabular}
\end{center}
\vglue 0.1cm

\vglue 0.1cm
\rightskip=3pc
\leftskip=3pc
{\tenrm \baselineskip=12pt
\noindent
Table 10. Errors for the normalized observables defined
at LHC implied by the NLC errors in Table 9.}
\vglue 0.1cm

\rightskip=0pc
\leftskip=0pc
\vglue 0.8cm
{\elevenbf\noindent Acknowledgements \hfil}
\vglue 0.3cm

It is a pleasure to thank M. Cveti\v c and P. Langacker
for their collaboration on the elaboration of this
review and
the organizers of the School for their warm hospitality.

\eject

{\elevenbf\noindent References \hfil}
\vglue 0.3cm

\end{document}